\documentclass[11pt]{article}
\usepackage{moriond,epsfig}

\def\be{\begin{equation}}
\def\ee{\end{equation}}
\def\bea{\begin{eqnarray}}
\def\eea{\end{eqnarray}}

\begin{document}
\vspace*{4cm}
\title{ B PHYSICS PROSPECTS AT LHC}
\author{ MARTA CALVI}
\address{Dipartimento di Fisica Universit\`a di Milano Bicocca 
and INFN Sezione di Milano}

\maketitle\abstracts{Future experiments at the LHC will have the opportunity
to pursue an extensive program on $B$ Physics and CP violation.
The expected performances will be presented here.}

\section{Introduction}
First pp collisions are foreseen in the Large Hadron Collider at Cern
in summer 2007.
At that time several precise measurements will be available from
B-Factories and Tevatron experiments to test the 
CKM paradigm of flavour structure and CP violation.
However New Physics could be hidden in B decays, specially those 
involving box and penguin diagrams.
On the other hand, if New Physics will be found at LHC in direct searches, 
B physics measurements will have to sort out the corresponding 
flavour structure.

The B Physics program at LHC is rich. It will include
precise measurement of $B^0_s\bar{B^0_s}$ mixing (mass difference,
width difference and phase),
precise measurements of the angle $\gamma$ (including from processes 
only at tree level),
several measurements of other CP phases in different channels for 
over-constraining the Unitarity Triangles, search for New Physics 
effects appearing
in rare exclusive and inclusive B decays,
studies of b-baryons and B$_c$ physics and also studies of b production.

B physics at LHC has the great advantage of a high $b\bar{b}$
cross section ($\sigma_{bb} \sim 500 \mu$b), 
several orders of magnitude higher than at the 
$\Upsilon(4S)$, and of the production of all species of $b$-hadrons, 
including $B_s$, $b$-baryons and $B_c$. 
The challenge in the analysis is related to 
the presence of the underlying event, to the high particle multiplicity 
and to the high rate of background events
(the inelastic cross section is $\sim 80$ mb ).
 These features demand to the experiments an excellent trigger capability, with
good efficiency also on fully hadronic decay modes of $b$-hadrons,
excellent tracking and vertexing performance, allowing for high
mass resolution and proper time resolution, and 
excellent particle identification to separate exclusive decays.

\section{LHC experiments}
The LHC will collide protons at 14 TeV with a bunch crossing rate of 40MHz.

Two experiments, ATLAS and CMS, are omni-purpose and 
optimized for discovery physics.
Their B physics program is mainly pursued in the first years of running, 
when the LHC luminosity is expected to be 
1-2$\times 10^{33}$cm$^{-2}$s$^{-1}$. 
In the subsequent years at high luminosity ($10^{34}$cm$^{-2}$s$^{-1}$),
when several pp collisions per bunch crossing will pile up, only search for
very rare B decays with clear signatures will be performed.
Reaches in B Physics will depend on the chosen trigger strategy and 
allocated bandwidth. 
B events will be mainly triggered by high $p_T$ muons or di-muon triggers.
CMS also exploits on-line tracking for the selection of exclusive B events 
at High Level Triggel ~( \cite{CMS}), 
while ATLAS foresees a flexible trigger strategy with the 
progressive addition of other triggers~( \cite{ATLAS}).

\begin{figure}[t]
 \vspace{9.0cm}
 \includegraphics{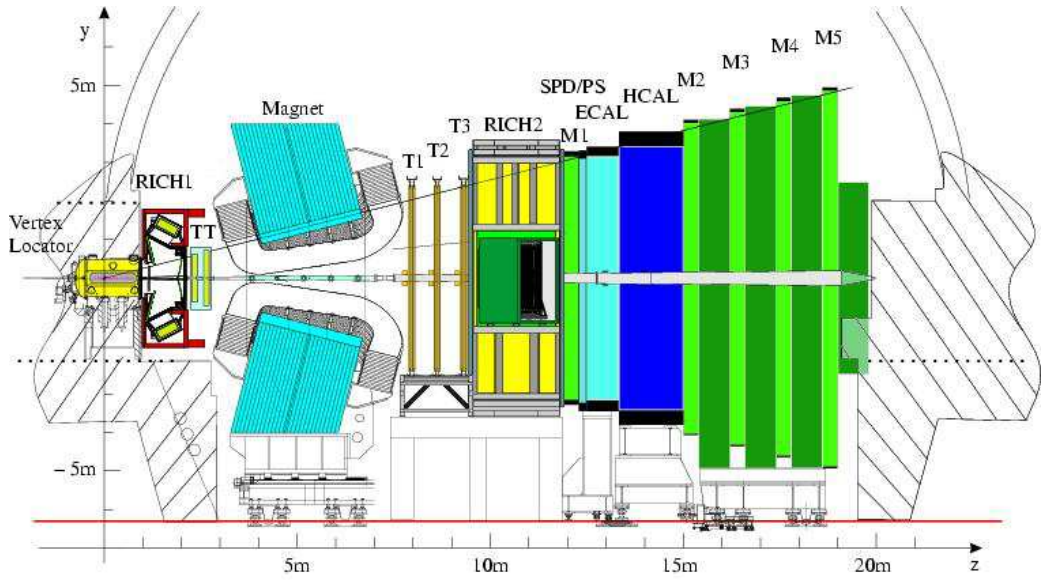}
 \caption{\it 
  A schematic view of the LHCb detector.
     \label{LHCb_scheme} }
\end{figure}

LHCb is the LHC experiment dedicated to B physics.
It will locally tune the luminosity, by de-focusing the beams, to  
2$\times 10^{32}$cm$^{-2}$s$^{-1}$, in order to limit pile up of pp
interactions.
Taking the nominal year period as $10^7$s, an integrated luminosity of 
2 fb$^{-1}$ per year is expected, corresponding to $10^{12}$ $b \bar{b}$ 
events/year.
LHCb is a single-arm forward detector in the polar region 10-300 mrad,
 with good acceptance for
 $b$ events due to the forward peaked production of $b$-hadrons at LHC.
A schematic view of the LHCb detector
is shown in Figure~\ref{LHCb_scheme}. A description of the detector 
and its performances can be found in~( \cite{LHCbLight}).
The LHCb trigger is operating in three stages. The Level-0 reduces the rate 
to 1 MHz requiring the presence of leptons or photons or hadrons with high
p$_T$ while the Level-1 selects on high impact parameter, high p$_T$ tracks.
The High Level Trigger is a software trigger using the full information on the
event. Its output contains 200 Hz of exclusive B candidates and 
about 1.8 KHz of inclusive channels to be used also for calibration purposes
and systematic studies.

\section{Prospects on $B^0_s\bar{B^0_s}$ mixing measurements }

\subsection{Measurements of $\Delta m_s$}
The  $B^0_s\bar{B^0_s}$  oscillation has been proven to be
too fast to be resolved at LEP and SLC experiments. The best limit today is
$\Delta m_s>14.5$ ps$^{-1}$ at 95\%CL.
The Tevatron is at present the only available source of $B^0_s$ mesons and 
CDF and D\O\  have the chance to find a mixing signal in the coming years.
The measurement requires best performances in the event reconstruction and
purity, proper time resolution and flavour tagging.

\begin{figure}[t]
 \vspace{9.0cm}
 \includegraphics{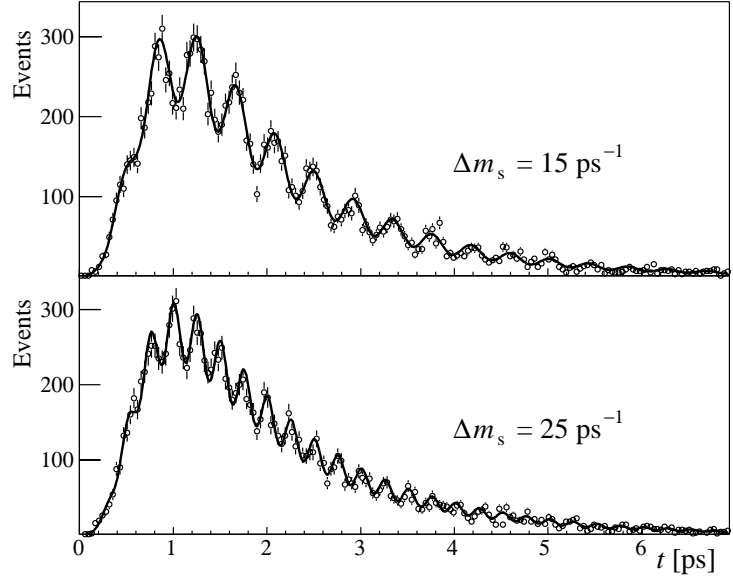}
 \caption{\it 
  Expected proper time distribution of simulated  
  $B^0_s \rightarrow D_s \pi$ candidates that have been flavour tagged as 
  having not oscillated, for two different values of $\Delta m_s$. 
  Points corresponds to one year of data-taking in LHCb.
  \label{fig_oscill} }
\end{figure}

But the definitive answer on $B^0_s\bar{B^0_s}$ mixing may come from LHC.
The best channel for these studies is $B^0_s \rightarrow D_s^- \pi^+$.
Results of LHCb full simulation indicate a proper time resolution of 
$\sigma_\tau \simeq 40$ fs 
and an annual yield of 80.000 events with a signal over background ratio 
of about 3. The effective efficiency for flavour tagging is estimated
to be about 7\%.
The expected proper time distribution of tagged events is shown in
Figure~\ref{fig_oscill} for two different values of $\Delta m_s$.
In one year of data-taking a 5$\sigma$ observation of oscillation 
 is expected if $\Delta m_s<68$ ps$^{-1}$.
Once observed, the precision 
to measure  $\Delta m_s$ is $\sim 0.01$ ps$^{-1}$ . 

ATLAS will also make a $5\sigma$ observation of oscillations if 
$\Delta m_s<22$ ps$^{-1}$, in 10 fb$^{-1}$. 
Most recent expectation of CMS is lower, due to restriction to the 
trigger bandwidth allocated to this channel.

\subsection{$\phi_s$ and $\Delta \Gamma _s / \Gamma _s $ measurements }

The phase $\phi_s$ of $B^0_s \bar{B^0_s}$ mixing is expected to be very 
small in the Standard Model
$\phi_s=-2 \chi= -2\lambda^2 \eta \simeq -0.04$ resulting in a high sensitivity
to possible New Physics contributions in  $b\rightarrow s$ transitions.
Hints of New Physics could also be found in the measurement of 
the decay width difference between the two CP eigenstates
$\Delta \Gamma _s =  \Gamma( B_L) - \Gamma(B_H)$.
In the Standard Model $\Delta \Gamma_s$ is expected to be of the order of 10\%.
Both quantities can be measured using $B^0_s \rightarrow J/\psi \phi$ decays
$(J/\psi \rightarrow \mu\mu,\phi \rightarrow KK)$.
In a decay to two vector mesons three distinct amplitudes contribute:
two CP even and one CP odd. The CP components can be disentangled 
on a statistical basis taking into account the distribution of the so-called 
transversity angle $\theta_{tr}$, defined as the angle between the positive 
lepton and the $\phi$ decay plane, in the $J/\psi$ rest frame.
The physics parameters can be extracted from a simultaneous fit to the proper 
time, $cos(\theta_{tr})$ and $\Delta m_s$ distributions of tagged events.
In one year of data-taking LHCb expects to collect 100.000  
$J/\psi(\mu\mu) \phi$ decays and to obtain
a precision on $sin(\phi_s)$ of about 0.06 
and precision on $\Delta \Gamma_s/ \Gamma_s$ of about 0.018 (for 
$\Delta m_s$=20 ps$^{-1}$).
The sensitivity will be increased adding 
 $B^0_s \rightarrow J/\psi\eta$ events, which are pure CP eigenstates.
About 7000 events per year are expected in this channel.

CMS and ATLAS, with 30 fb$^{-1}$, expect a sensitivity on $sin(\phi_s)$ of 
0.03-0.04 and a sensitivity on $\Delta \Gamma_s/ \Gamma_s$ of 0.015-0.012, 
respectively.

\section{Measurements of $\gamma$}

\subsection{$\gamma$ measurements from $B_s\rightarrow D_sK$ decays }

\begin{figure}[t]
 \vspace{9.0cm}
 \includegraphics{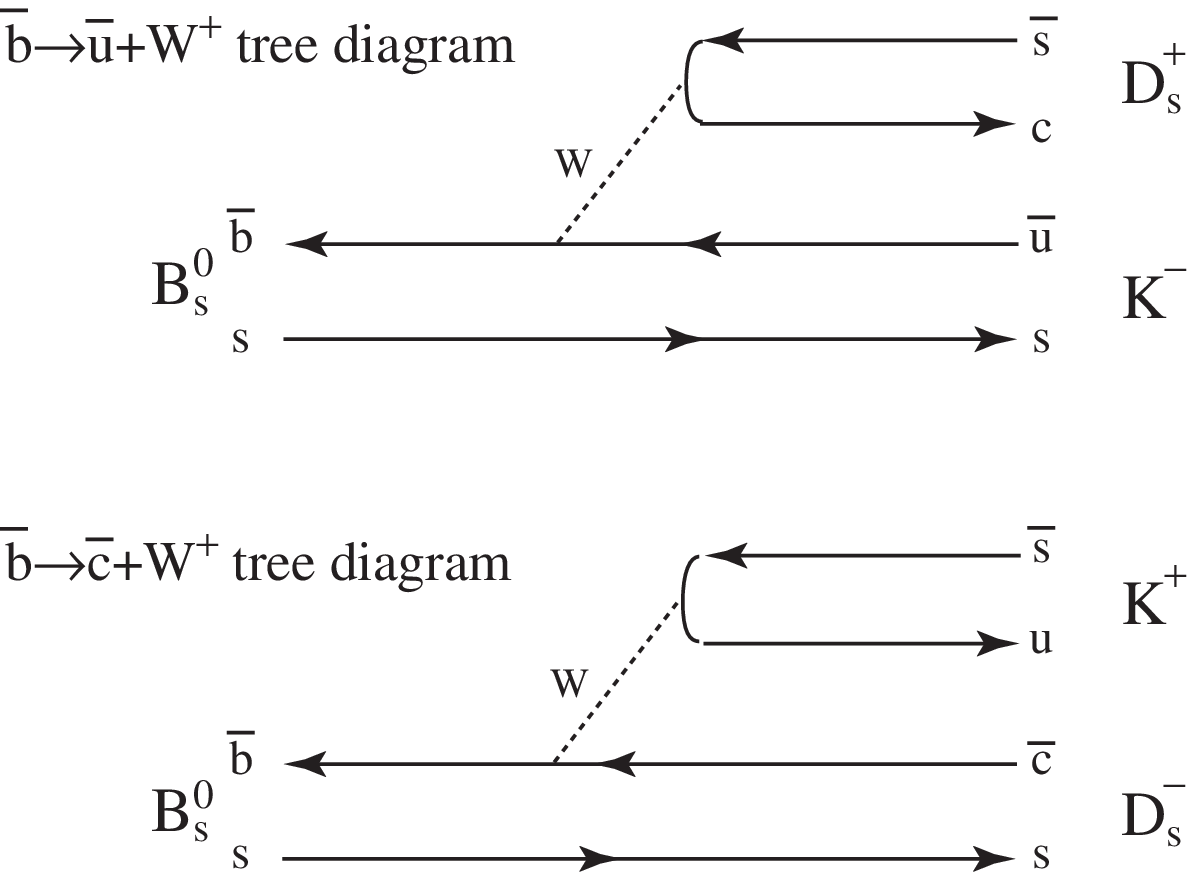}
 \includegraphics{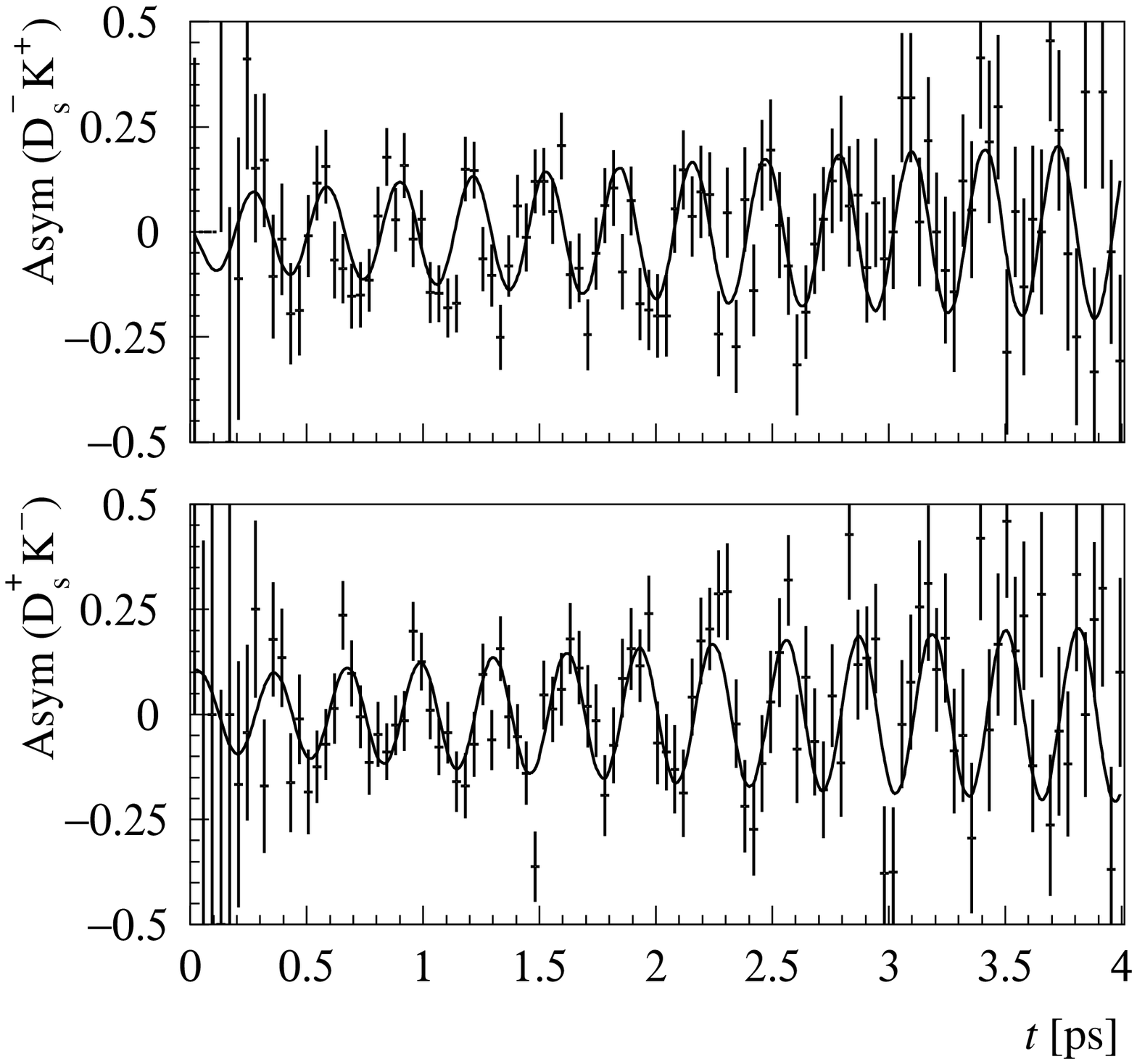}
 \caption{\it 
  Left: The $B^0_s\rightarrow D_s^+ K^-$ and  $B^0_s\rightarrow D_s^- K^+$
   tree diagrams.
  Right: Time-dependent  $B^0_s \bar{B^0_s}$ asymmetry of simulated $D_s^-K^+$
  (top) and $D_s^+K^-$ (bottom) candidates, for $\Delta m_s=20$ ps$^{-1}$.
  Points correspond to 5 years of data-taking.
  \label{fig_dsk} }
\end{figure}

$B^0_s\rightarrow D_s^\pm K^\mp$ and  $\bar{B^0_s}\rightarrow D_s^\mp K^\pm$ 
decays can proceed through the tree diagrams shown in 
Figure~\ref{fig_dsk}, which can interfere via mixing.
From the measurement of the time-dependent decay asymmetries
the phase $\gamma+\phi_s$ can be extracted, together with a strong phase.
If $\phi_s$ has been determined otherwise, $\gamma$ can be extracted,
with little theoretical uncertainty, and insensitive to New Physics. 

\begin{figure}[t]
 \vspace{7.0cm}
 \includegraphics{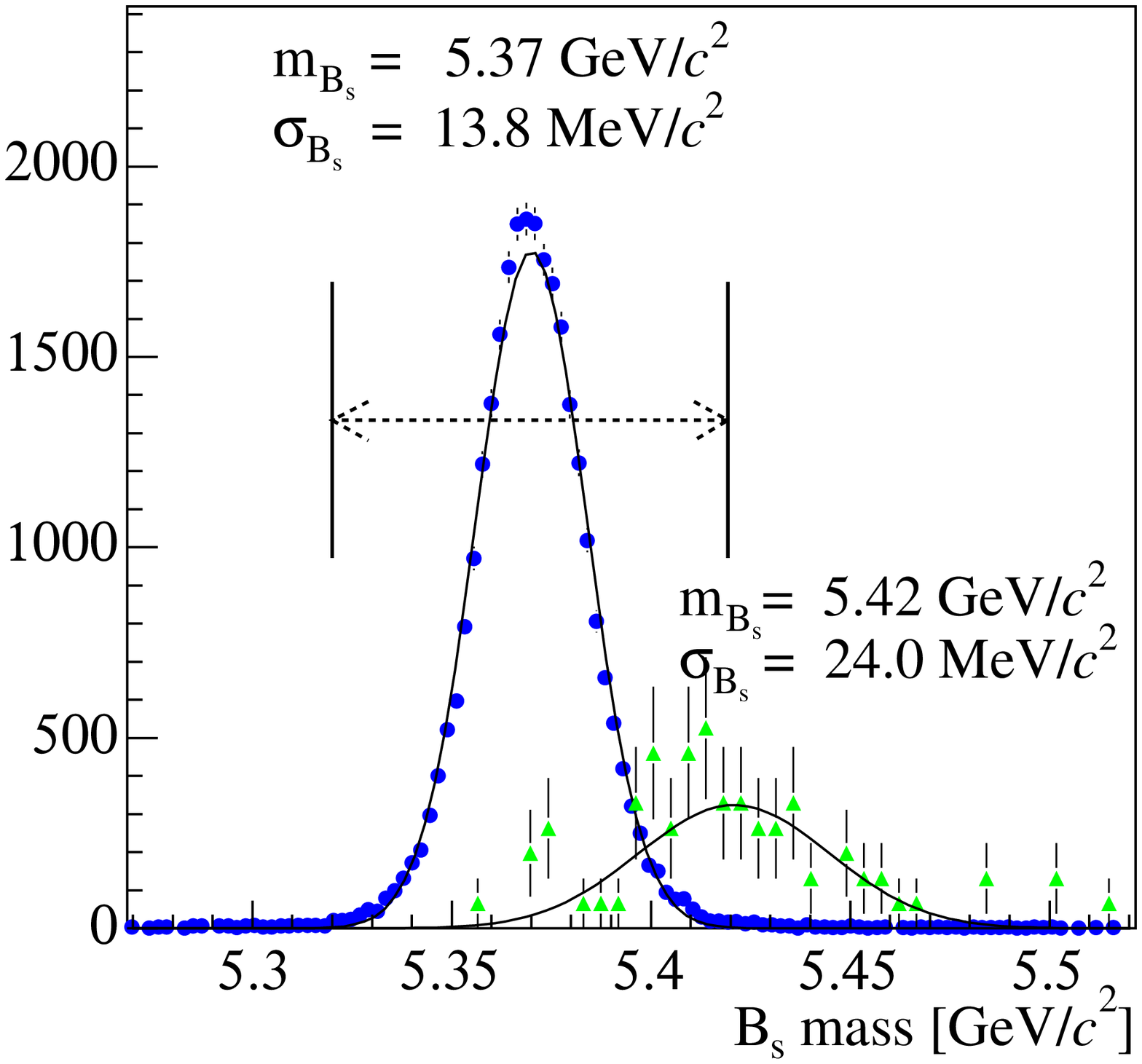}
 \includegraphics{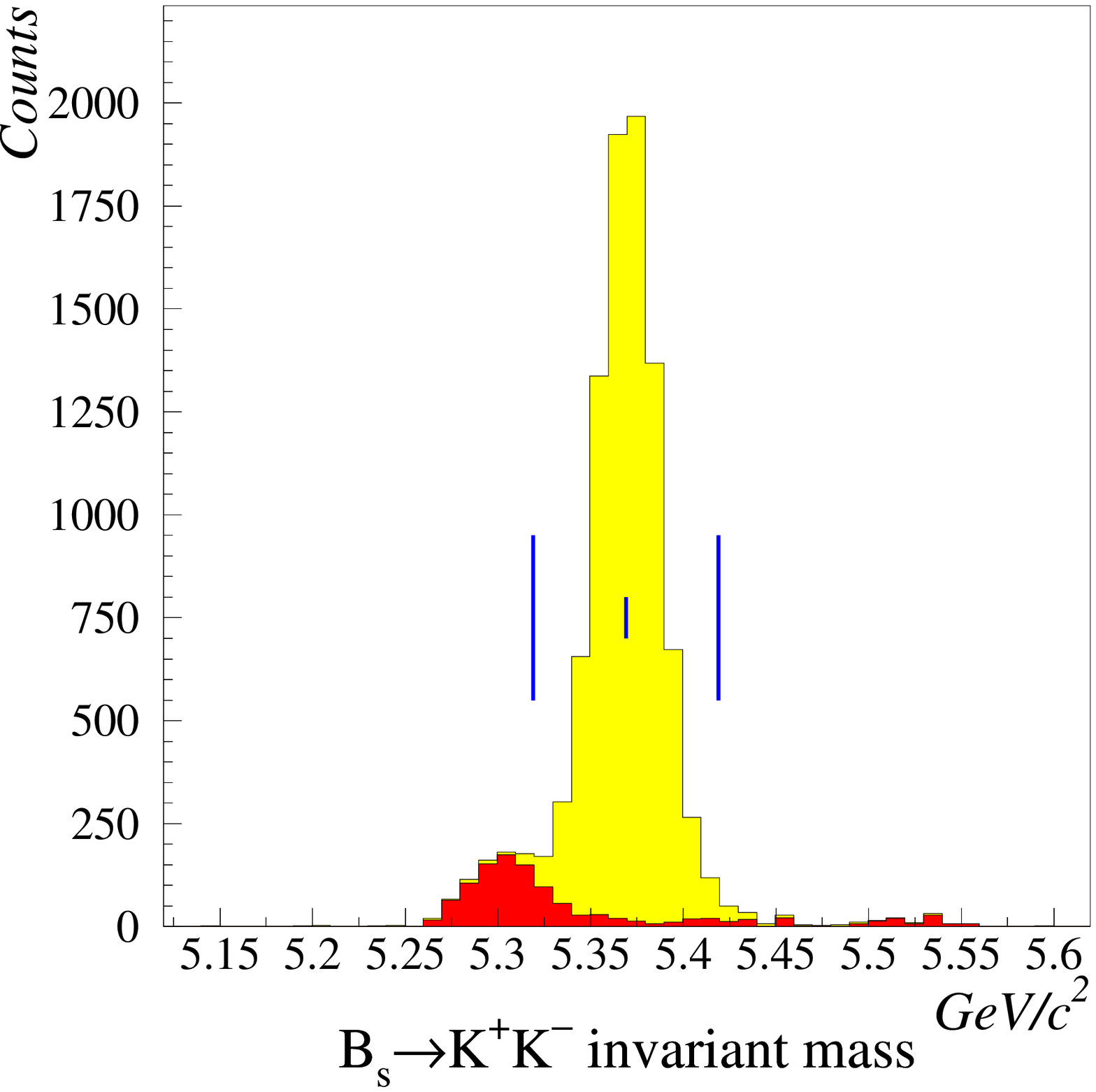}
 \caption{\it 
  Left: The  $B^0_s$ mass distribution of $B^0_s\rightarrow D_sK$ decays,
  as determined from the full LHCb Monte Carlo. 
  Also shown is the mass distribution of $B^0_s\rightarrow D_s\pi $ decays 
  that have been reconstructed as $B^0_s\rightarrow D_sK$ decays. 
  The residual contamination is about 10\%.
  The histograms are correctly normalized to compensate for the different 
  branching ratios.
  Right: invariant mass distribution of selected  $B^0_s\rightarrow K^+K^-$
  candidates in LHCb . The light shaded histogram shows the signal and the dark
  one represents the background from $B^0\rightarrow \pi^+\pi^-,K^+\pi^-$;
  $B^0_s\rightarrow \pi^+K^-$; $\Lambda_b\rightarrow pK^-,p\pi^-$.
  \label{fig_massplot} }
\end{figure}

Strong particle identification capabilities are required to 
separate $B_s\rightarrow D_sK$ decays from $B_s\rightarrow D_s\pi$ 
background having $\sim$12 times larger branching fraction.
The performance of the LHCb RICH detectors will be fully adequate, as it
is shown in Figure~\ref{fig_massplot}. 
Monte Carlo studies have shown that 5400 $D_s^\mp K^\pm$ events 
will be collected in one year of data-taking, with S/B ratio, estimated from 
$b\bar{b}$ events, larger than 1. Hereafter only limits on S/B are quoted,
due to the limited statistics with which the value of B has been determined.
The $D_s^\mp K^\pm$  asymmetries are shown in Figure~\ref{fig_dsk}.
A sensitivity of $\sigma_\gamma=14$ degrees can be obtained
if $\Delta m_s$=20 ps$^{-1}$.

\subsection{$\gamma$ measurements from $B^0\rightarrow D^0 K^{*0}$ decays }
\begin{figure}[t]
 \vspace{7.0cm}
 \includegraphics{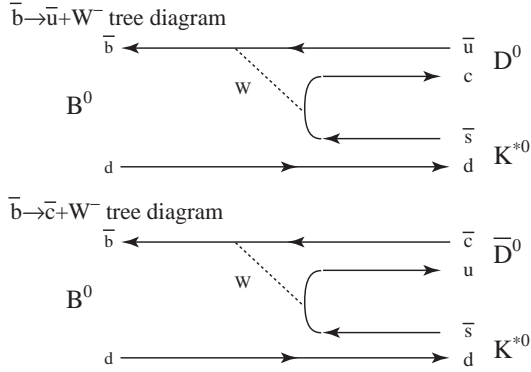}
 \caption{\it 
  The $B^0\rightarrow D^0 K^{*0}$ and  $B^0\rightarrow \bar{D^0} K^{*0}$
   tree diagrams.
  \label{fig_dkstar-diagram} }
\end{figure} 

A theoretically clean determination of the angle $\gamma$ can be performed
using $B^0\rightarrow D^0 K^{*0}$ decays. The tree diagrams of the decay are
shown in Figure~\ref{fig_dkstar-diagram}.
The method, described in~( \cite{Dunietz}), is based on the measurement 
of six time-integrated decay rates for
$B^0_d\rightarrow D^0 K^{*0},\bar{D^0}K^{*0},D_{CP}K^{*0}$
 and their CP conjugates; the decays are self-tagged through 
$K^{*0}\rightarrow K^+\pi^-$, while the CP auto-states 
$D_{CP}$ can be reconstructed in $K^+K^-$ and $\pi^+\pi^-$ modes.
This method is similar to the analysis of $B^\pm\rightarrow D^0 K^\pm$ decays, 
already performed at the B-Factories,
but has the advantage of using two colour-suppressed diagrams with 
an expected ratio of amplitudes $r=|A(B^0\rightarrow D^0 K^{*0})| / 
|A(B^0\rightarrow \bar{D^0} K^{*0})|\sim 0.4$. 

LHCb in one year of data taking expects to collect a total of about 4.000 
signal events leading to a sensitivity on $\gamma$ of 
$\sigma_\gamma \sim 8$ degrees.

\subsection{$\gamma$ measurements from
$B^0_d\rightarrow \pi^+\pi^-$ and  $B^0_s\rightarrow K^+K^-$ decays}

Several strategies have been proposed~( \cite{CKM})  to extract
informations on the angle $\gamma$ from two body charmless decays of
$B$ mesons, some of them make use of assumptions on dynamics or on 
U-spin flavour symmetry. 

RICH detectors allow to separate the $K/\pi$ 
channels with high efficiency and purity, as shown in 
Figure~\ref{fig_massplot}.
LHCb in one year of data taking expects to collect 
26.000 $B^0_d\rightarrow \pi^+\pi^-$, 
37.000 $B^0_s\rightarrow K^+K^-$ and
135.000 $B^0_d\rightarrow K^+\pi^-$ decays,  
with mass resolution $\sigma(M_B)\sim17$ MeV and a proper time resolution of
$\sigma_\tau \sim 30$ fs. 
The two time dependent CP asymmetries
$$A_{CP}({B^0_d\rightarrow \pi^+\pi^-})(t)=A_{CP}^{dir,\pi\pi}cos(\Delta m_d t)
                                     +A_{CP}^{mix,\pi\pi}sin(\Delta m_d t)$$
$$A_{CP}({B^0_s\rightarrow K^+K^-})(t)=A_{CP}^{dir,KK}cos(\Delta m_s t)
                                     +A_{CP}^{mix,KK}sin(\Delta m_s t)$$
will be used to fit the four CP asymmetries, which will be extracted 
with a  precision of about 6\%.
Following the method suggested in~( \cite{Fleisher}), 
U-spin symmetry can be exploited to constrain the penguin to tree 
ratios in the two decays to be the same. Assuming the knowledge 
of the mixing phases $\phi_d$ and $\phi_s$ from previous measurements, 
the $\gamma$ angle can be extracted with a precision 
$\sigma_\gamma \sim 5$ degrees in one year of data-taking.
Additional measurements can be used to test the uncertainty 
related to the U-spin assumptions.

\section{$\alpha$ measurements from $B^0\rightarrow\rho\pi$ decays }
\begin{figure}[t]
 \vspace{7.0cm}
 \includegraphics{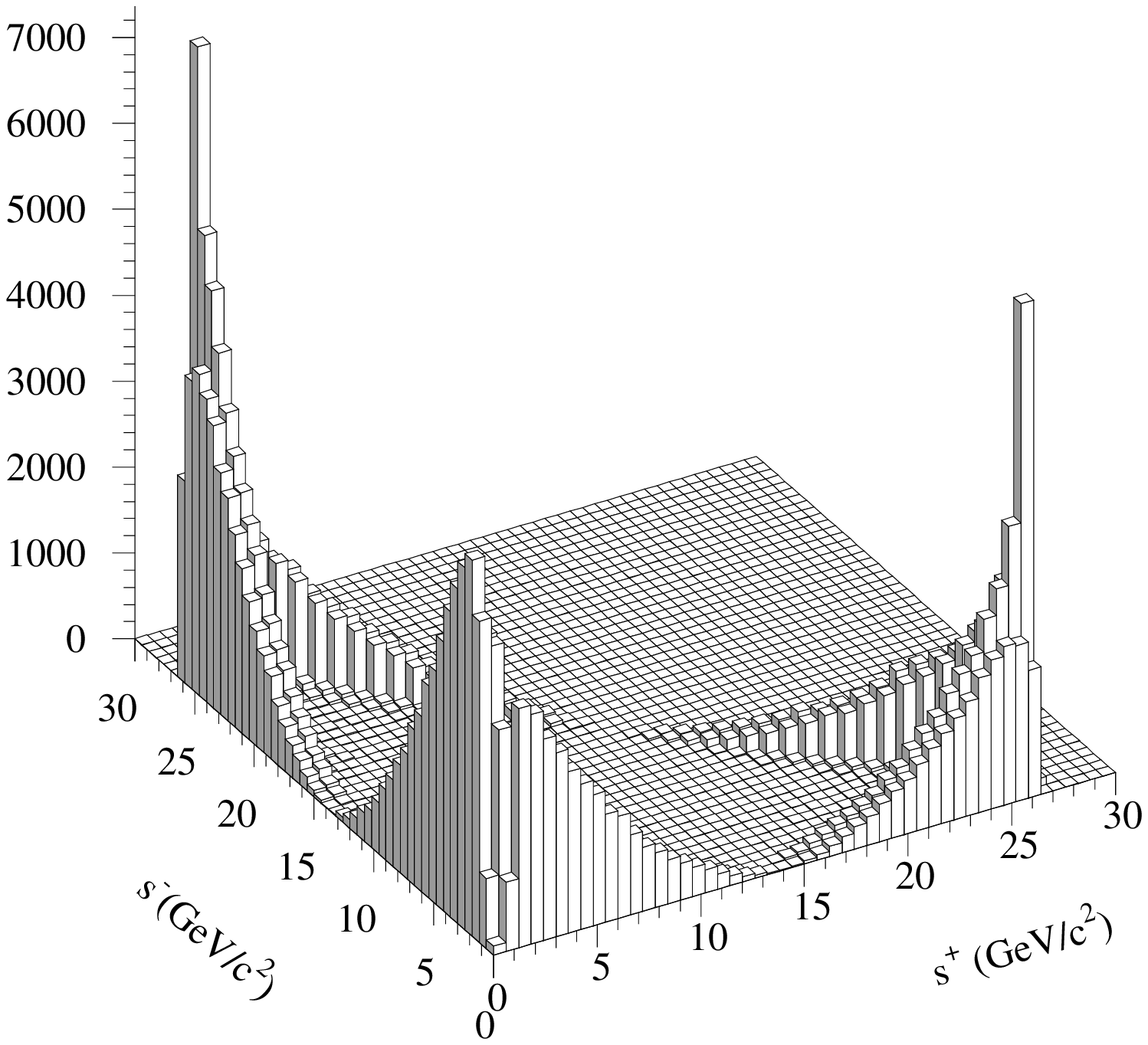}
 \includegraphics{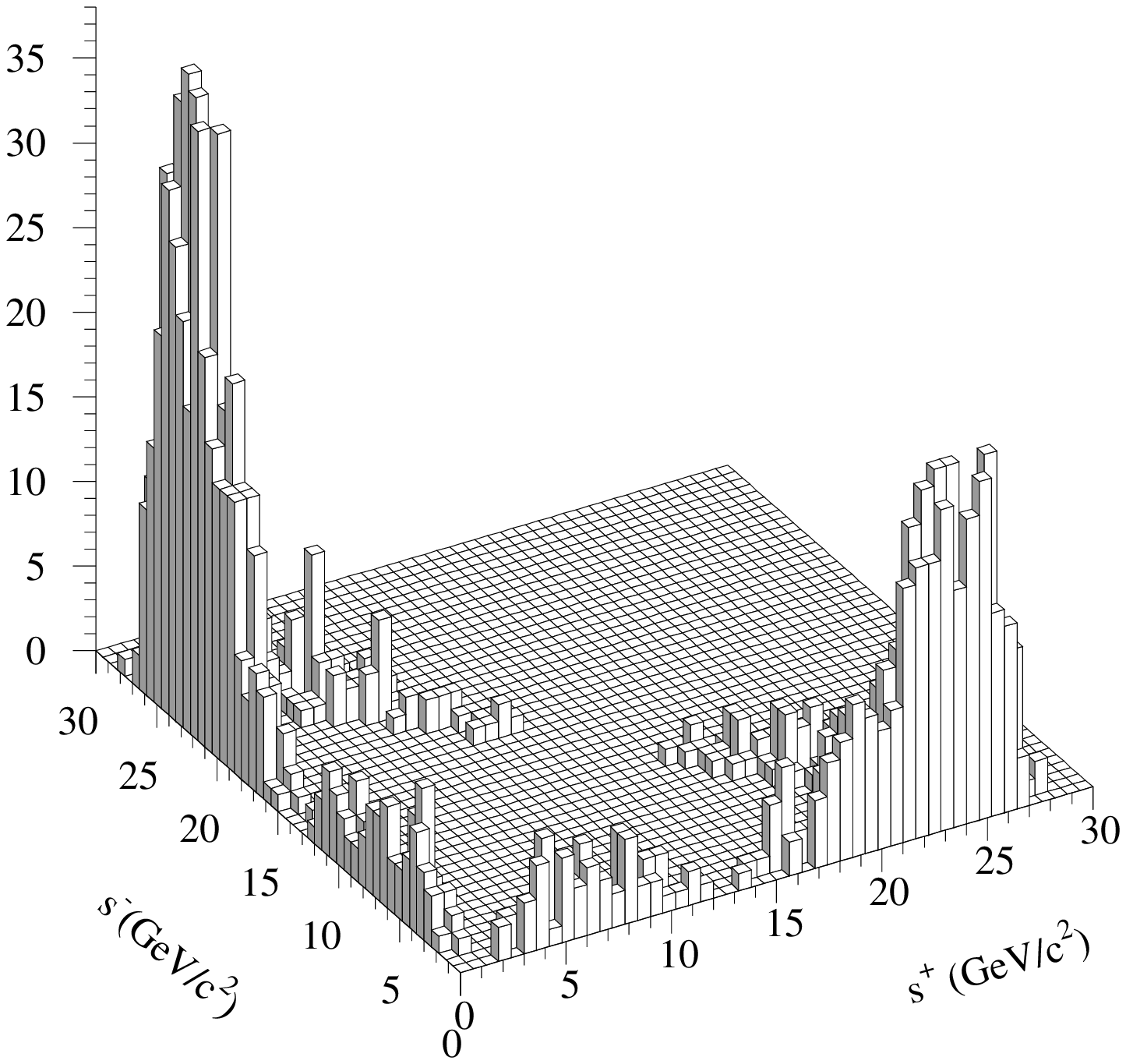}
 \caption{\it Dalitz distribution of $B^0\rightarrow\pi^+\pi^-\pi^0 $ events
in LHCb. The left plot correspond to all Monte Carlo events, the 
right plot to the reconstructed and selected events. $s^+=M^2(\pi^0\pi^+)$, 
$s^-=M^2(\pi^0\pi^-)$. 
     \label{LHCb_Dalitz} }
\end{figure}

A time dependent Dalitz plot analysis of the three body decay
$B^0\rightarrow\rho\pi\rightarrow\pi^+\pi^-\pi^0$ 
allows a clean extraction of the angle 
$\alpha=\pi-\beta-\gamma $, as suggested in~( \cite{Snyder-Quinn}).
LHCb expects to reconstruct 14000 decays per year (with S/B$>1.3$)
in this channel. 
The Dalitz distribution is shown in Figure~\ref{LHCb_Dalitz}. 
An 11-parameter fit has been used to get an independent measurement 
of tree and penguin parameters,
taking into account resonant and non resonant background sources.
A sensitivity $\sigma_\alpha<$10 degrees can be obtained in one year of
data-taking.

\section{$B^0_d\rightarrow K^{*0}\mu^+\mu^-$ }

$B^0_d\rightarrow K^{*0}\mu^+\mu^-$ is a rare decay with
branching fraction of the order of $10^{-6}$ which has a clear 
experimental signature.
The  forward-backward asymmetry is defined as
$$A_{FB}(\mbox{\^s})=(\int_0^1 dcos\theta - \int_{-1}^0 dcos\theta) 
{{d\Gamma^2}\over{d\mbox{\^s} dcos\theta}}$$
where $\theta$ is the angle between the $\mu^+$ and the $K^{*0}$ in the
di-muon rest frame, and \^s=$( m_{\mu^+\mu^-}/ m_B)^2$.
The forward-backward asymmetry is a sensitive probe of New Physics. 
In the Standard Model the value of \^s
for which $A_{FB}$(\^s) is zero can be calculated with a 5\% precision.
Models with non-standard values of Wilson coefficients
$C_7,C_9,C_{10}$ predict $A_{FB}$(\^s) of opposite sign or without zero point.

LHCb will select 4400 decays per year (with S/B$>0.4$), 
this allows a determination of branching fractions and CP asymmetries with a 
precision of few percent.
Using a toy Monte Carlo to determine the sensitivity in the forward-backward
asymmetry measurement, including background subtraction, an uncertainty
of 0.06 on the location of \^s$_0$ is found, in 1 year of data-taking.

ATLAS will also collect about 2000 events of $B^0_d\rightarrow K^{*0}\mu^+\mu^-$, 
with S/B$>7$ in 30 fb$^{-1}$.

\section{$B_s\rightarrow \mu^+\mu^-$ }
$B_s\rightarrow \mu^+\mu^-$  is a rare decay involving flavour changing
neutral currents whose branching ratio is estimated to be
$BR (B_s\rightarrow \mu^+\mu^-)=(3.5\pm0.1)\times 10^{-9}$ in the 
Standard Model~( \cite{Ali}). 
In various supersymmetric extensions of the Standard model 
it can be enhanced by one to three orders of magnitude, 
being $BR \sim (tan\beta)^6$, for large $tan\beta$. 
The best upper limit on the branching ratio at present come from  
experiments at Tevatron, and reachs few$\times 10^{-7}$ at 95\% CL.

In the SM context, LHCb expects to select 17 events per year, with a 
resolution on the B$_s$ mass of 18 MeV/c$^2$. The
background determination is still incomplete and require additional
Monte Carlo statistics. No events were selected in the $10^7$ $b\bar b$
event sample used so far.

CMS has studied a selection at the High Level Trigger, giving
$B_s\rightarrow \mu^+\mu^-$ candidates 
with a rate smaller than 1.7 Hz, and a resolution  on the B$_s$ mass
of 74 MeV/c$^2$. In 10 fb$^{-1}$ 47 signal events are selected. 
With a refined selection at the offline level 7 signal
events are expected to be retained, with less than 1 background event.

For this search ATLAS and CMS will also exploit the high luminosity runs. 
In 100 fb$^{-1}$ (1 year at 10$^{34}$) 
    92 signal events are expected (with 660 of background) 
and 26 signal events (with $\sim 6$ of background) respectively.
The different levels of background can be attributed to different 
vertex reconstruction and selection, however an update on these estimations
is expected.

In conclusion there are good prospects of significant measurement 
in this channel, even for the  SM value of the branching ratios.

\section{Conclusion}
In the coming years CP asymmetries will be measured at LHC
using several $B^0_d$ and $B^0_s$ mesons and $b$-baryons decay channels.
Very rare decays will also be studied, thanks to the high  
$b\bar{b}$ cross section available.
This program is complementary to the B-Factories one and will
allow to complete and improve the available results and possibly to reveal 
first signals of new Physics.


\end{document}